\documentclass[trackchanges,preprint]{aastex62}
\usepackage{gensymb}
\usepackage{enumerate}

\received{April 10, 2018}
\revised{August 14, 2018}
\accepted{August 19, 2018}

\submitjournal{\apj}
\shorttitle{New Galactic SNRs}
\shortauthors{Dokara et al.}

\turnoffediting

\begin{document}

\title{Confirmation Of Two Galactic Supernova Remnant Candidates Discovered By THOR}

\correspondingauthor{Rohit Dokara}
\email{dokararohit@iisc.ac.in}

\author[0000-0002-1971-6725]{Rohit Dokara}
\affiliation{Department of Physics, Indian Institute of Science, Bengaluru 560012, India}

\author[0000-0001-9829-7727]{Nirupam Roy}
\affiliation{Department of Physics, Indian Institute of Science, Bengaluru 560012, India}

\author[0000-0002-1700-090X]{Henrik Beuther}
\affiliation{Max Planck Institute for Astronomy, K\"{o}nigstuhl 17, 69117, Heidelberg, Germany}

\author[0000-0001-8800-1793]{L. D. Anderson}
\affiliation{Department of Physics and Astronomy, West Virginia University, Morgantown WV 26506, USA}
\affiliation{Adjunct Astronomer at the Green Bank Observatory, P.O. Box 2, Green Bank WV 24944, USA}
\affiliation{Center for Gravitational Waves and Cosmology, West Virginia University, Chestnut Ridge Research Building, Morgantown, WV 26505, USA}

\author{Michael Rugel}
\affiliation{Max Planck Institute for Astronomy, K\"{o}nigstuhl 17, 69117, Heidelberg, Germany}

\author[0000-0003-2623-2064]{Jeroen Stil}
\affiliation{Department of Physics and Astronomy, University of Calgary, 2500 University Drive NW, Calgary AB, T2N 1N4, Canada}

\author{Yuan Wang}
\affiliation{Max Planck Institute for Astronomy, K\"{o}nigstuhl 17, 69117, Heidelberg, Germany}

\author[0000-0002-0294-4465]{Juan D. Soler}
\affiliation{Max Planck Institute for Astronomy, K\"{o}nigstuhl 17, 69117, Heidelberg, Germany}

\author{Russel Shanahan}
\affiliation{Department of Physics and Astronomy, University of Calgary, 2500 University Drive NW, Calgary AB, T2N 1N4, Canada}

\begin{abstract}

\citet{2017A&A...605A..58A} identified seventy six candidate supernova remnants
(SNRs) using data from The HI, OH, Recombination line survey of the Milky Way
(THOR). The spectral index and polarization \edit2{properties } can help
distinguish between SNRs and H\begin{small}II\end{small}regions, which are
often confused. We confirm \edit1{two } SNR candidates using spectral index data
and morphology. However, we observe that the fractional linear polarization
cannot distinguish between SNRs and H\begin{small}II\end{small}regions,
\edit1{likely due to contamination by diffuse Galactic synchrotron
emission}. We also comment on the association of SNR candidates with pulsars
\edit1{through geometric and age considerations.}

\end{abstract}

\keywords{ISM: H\begin{small}II\end{small}regions --- ISM: Supernova remnants --- Polarization}

\section{Introduction}

The list of nearly 300 Galactic Supernova Remnants (SNRs) compiled by
\citet{2014BASI...42...47G} is thought to be incomplete because it was
estimated that there must be upwards of 1000 SNRs in the Milky Way
\edit1{\citep{1991ApJ...378...93L,1994ApJS...92..487T}}. Even including recent
TeV $\gamma$-ray SNRs detected by the H.E.S.S. collaboration does not increase
this number greatly \citep{2017AIPC.1792d0030G}. Though the lack of detections
at radio and X-ray wavelengths supports the arguments of
\citet{2006MNRAS.371.1975Y} that there could be several SNRs with no radio,
optical, ultraviolet or X-ray emissions, \citet{2006ApJ...639L..25B} have shown
that the deficiency is primarily due to the lack of sensitivity to observe low
surface brightness SNRs and due to low angular resolution that prevents the
detection of small angular size SNRs.

Galactic SNRs are routinely identified in radio wavelengths. The emission (or
the lack of it) at different wavelengths depends on intrinsic factors such as
the progenitor's history and also on external conditions such as the properties
of surrounding medium. H\begin{small}II\end{small}regions, which
are bright at radio wavelengths due to thermal emission, are frequently confused
with SNRs. SNRs have a significantly smaller ratio of flux at Mid Infrared (MIR)
wavelengths to flux at radio wavelengths than
H\begin{small}II\end{small}regions
\edit1{\citep{2001MNRAS.325..531C}}. This feature led to the proposal of 76
candidate SNRs by \citet{2017A&A...605A..58A} who have used radio continuum
data from The HI, OH, Recombination line survey of the Milky Way
\citep[THOR;][]{2016A&A...595A..32B} and the Karl G. Jansky Very Large Array
(VLA) 1.4 GHz Galactic Plane Survey \citep[VGPS;][]{2006AJ....132.1158S}, and
MIR wavelength data from Spitzer GLIMPSE, Spitzer MIPSGAL and WISE surveys.

We propose to confirm the \edit1{identification of the candidate SNRs} by
measuring the fractional linear polarization and spectral index of the total
\edit1{continuum} emission. We \edit1{measured these parameters for} known
SNRs and known H\begin{small}II\end{small}regions in the THOR
survey region ($|b| <$ 1.25\degree, 17.5\degree $< l <$ 67.4\degree;
\citealp[see][]{2016A&A...595A..32B}) \edit1{and compared them with the data
from candidate SNRs.} The list of known
H\begin{small}II\end{small}regions was taken from The WISE
Catalog of Galactic H\begin{small}II\end{small}regions
\citep{2014ApJS..212....1A} through an interactive website
\citep{2014AAS...22331201A}. We leave out candidate
H\begin{small}II\end{small}regions and use only known
H\begin{small}II\end{small}regions with sizes greater than $1'$,
so that the comparison with known and candidate SNRs is appropriate.

Pulsars near the SNR candidates with associations to high energy sources could
\edit1{be indicative of a positive identification. However, to ensure a clear
identification, distances to the SNR and pulsar should be compatible with age
and proper motion of the pulsar.} The Australia Telescope National Facility
(ATNF) pulsar catalog \citep{2005AJ....129.1993M} provides the list of
pulsars and their association with other sources.

H\begin{small}II\end{small}regions are expected to have no
linearly polarized emission at radio wavelengths because their emission is
thermal. They have flat radio spectra ($\alpha \approx 0$)\footnote{Spectral
index $\alpha$ is defined by $S_\nu \propto \nu^{\alpha}$, for a flux density
$S_\nu$ and a frequency $\nu$} for optically thin and $\alpha \gtrsim 0.5$ for
optically thick regions.

On the other hand, SNRs are strong synchrotron sources, which are highly
linearly polarized. For synchrotron emission in a uniform magnetic field,
fractional linear polarization is related to the spectral index by
\citep{2013tra..book.....W}:
\begin{equation}
  \Pi \equiv \frac{\sqrt{Q^2 + U^2}}{I} = \frac{3-3\alpha}{5-3\alpha}
\end{equation}
Synchrotron emission usually has $-2 < \alpha < 0$, so we expect fractional
linear polarizations of above 0.6. However, we rarely observe $\Pi > 0.25$.
This is due to the Faraday depolarization effect \citep{2007EAS....23..109F}.
Varying rotation measure within the resolution element causes different Faraday
rotations of the polarization vector, leading to reduced polarization fraction
being observed. As the polarization data from THOR is not fully processed yet
\citep{2016A&A...595A..32B}, we use the polarization data from the 1.4 GHz
Northern VLA Sky Survey \citep[NVSS;][]{1998AJ....115.1693C}.

We use the publicly available spectral index map presented by
\citet{2018MNRAS.474.5008D}, which is based on data from 150 MHz TIFR
GMRT\footnote{TIFR -- Tata Institute of Fundamental Research; GMRT --
Giant Metrewave Radio Telescope} Sky Survey (TGSS) \citep{2017A&A...598A..78I}
and the NVSS.

\edit2{RRLs are indicative of thermal processes, but the lines are often
broad and weak.} \edit1{\citet{2016A&A...595A..32B} have detected RRLs in THOR
survey only in some H\begin{small}II\end{small}regions. Hence, while a
detection of RRLs in THOR might imply thermal emission, a non-detection does
not imply non-thermal emission, and it cannot be used to confirm the SNR
candidates.}

\section{Data}

\subsection{List of known and candidate SNRs}

The catalog by \citet{2017yCat.7278....0G} contains the list of all SNRs
confirmed until 2016 along with their angular sizes. Of the 57 cataloged
known SNRs in the THOR survey region, some SNRs are not visible in the 1.4 GHz
THOR continuum data and some were classified as being
H\begin{small}II\end{small}regions by
\citet{2017A&A...605A..58A}. Such objects are excluded from calculations,
leaving us with 49 known SNRs. The list of SNR candidates is taken in its
entirety from the \citet{2017A&A...605A..58A} paper.
\citet{2017A&A...605A..58A} have combined \edit2{the THOR 1.4 GHz continuum
data (resolution $\sim 20''$)} with VGPS data, which is called ``THOR+VGPS''.
\edit2{The combined THOR+VGPS data have a spatial resolution of $25''$}. The
THOR survey was taken by the VLA in C configuration, whereas VGPS was taken
with the VLA in D configuration, and VGPS has single dish continuum data added
from Effelsberg telescope \citep{2006AJ....132.1158S}.

\subsection{Northern VLA Sky Survey (NVSS)}

The NVSS covered entire sky north of $-40\degree$ declination at 1.4 GHz and
its principal data products, maps of Stokes $I$, $Q$ and $U$, are provided
through a postage stamp
server\footnote{\url{https://www.cv.nrao.edu/nvss/postage.shtml}}
\citep{1998AJ....115.1693C}. The compact D and DnC configurations of the VLA
were used for the survey. Images were restored with a beam of $45''$. The
largest angular scale detectable is about $8'$. The RMS noise for Stokes $I$ is
$\approx$ 0.45 mJy beam$^{-1}$ and for Stokes $Q$, $U$ it is $\approx$ 0.29 mJy
beam$^{-1}$. The noise in all Stokes maps can be higher in the Galactic plane
by a factor of up to 1.5, depending on the observed region. We have measured the
source integrated flux density by retaining only those pixels above 2.5$\sigma$
level. To calculate polarized flux density, we apply this 2.5$\sigma$ mask to
the Stokes $Q$ and $U$ maps, and only those pixels above a 3$\sigma$ level
within this mask are retained. NVSS contains blank pixels in some regions due to
inadequate \edit2{sky} coverage and poor sensitivity. Four known SNRs, three SNR candidates
and 22 H\begin{small}II\end{small}regions contain a large portion
of blank pixels, in either Stokes $I$ or Stokes $Q$ \& $U$. These objects are
omitted in our calculations.

\subsection{TGSS-NVSS Spectral Index Map}

We utilize the \edit2{pixel by pixel} spectral index
map\footnote{\url{http://tgssadr.strw.leidenuniv.nl/doku.php?id=spidx\#spectral_index_map}}
created using 150 MHz TIFR GMRT Sky Survey (TGSS) and the 1.4 GHz NVSS, by
\citet{2018MNRAS.474.5008D}. The TGSS covers all sky north of $-53\degree$
declination at 150 MHz with a resolution of $25''$. \citet{2017A&A...598A..78I}
have made an alternative release of the data collected by the TGSS team. The
survey provides Stokes $I$ images. The median RMS noise is 3.5 mJy beam$^{-1}$.
However, in the Galactic plane, the RMS noise can be as high as 10 mJy
beam$^{-1}$. For the TGSS, flux densities are well recoverable up to the order
of a few arcminutes.

Similar to the NVSS, TGSS is solely interferometric data and lacks single dish
data. \citet{2016AcA....66...85W} have shown that for large structures (size
$> 16'$), \edit2{significant flux from extended regions is not detected in the
NVSS data because of the limited spatial frequency coverage of the
observations. Despite this problem}, the use of these data together to measure
spectral index is justified because the NVSS and TGSS have similar shortest
baselines and they are two of the most sensitive centimeter and meter
wavelength surveys with large sky coverage. \edit2{Furthermore, our criterion
to identify SNRs is based on steep negative spectral index. If the flux
undetected in the TGSS is indeed considerable, it would make the measured
spectral index larger. Hence, the problem of undetected flux in the TGSS ---
if it exists --- would make our criterion to detect SNRs even stronger.}

Though spectral indices can be derived from the THOR data between 1000 and 2000
MHz \citep{2016A&A...588A..97B}, the TGSS-NVSS spectral indices span an even
broader frequency range of a factor $\sim$10 between 150-1400 MHz. These should
be better suited for spectral index maps of SNR candidates with angular sizes of
several arcminutes. At these scales, the THOR data are affected by generally
stronger spatial filtering than NVSS due to the more extended array
configuration (VLA C-configuration).

\subsection{List of known pulsars and their associations}

The ATNF pulsar catalog contains all known spin-powered pulsars and
magnetars but excludes accretion powered systems. The data is publicly
available on the ATNF
website\footnote{\url{http://www.atnf.csiro.au/research/pulsar/psrcat/}}
\citep{2005AJ....129.1993M}. The list also provides distances, frequencies and
known associations of pulsars to \edit2{SNRs}, pulsar wind nebulae (PWNe), X-ray
and gamma-ray sources where available.

\section{Results and Discussion}

For the three samples (H\begin{small}II\end{small}regions, known
and candidate SNRs), we have measured linearly polarized flux density, total
flux density and fractional linear polarization at 1.4 GHz from the NVSS, and
the 150-1400 MHz spectral index. A significant number of
H\begin{small}II\end{small}regions overlap with known and
candidate SNRs. For instance, there are multiple bright
H\begin{small}II\end{small}regions in the shell type SNRs
G23.3$-$0.3 and G32.8$-$0.1. These introduce an
uncertainty in both polarization and spectral index calculations. Unrelated
sources such as Active Galactic Nuclei (AGN) also might affect the
measurement of polarization and spectral index. \edit1{AGNs comprise most of the
unrelated background sources. They are compact and have $\alpha\approx-1$.}

\subsection{Spectral Index}
\label{subsec:spidx}

\begin{figure*}
  \centering
  \plotone{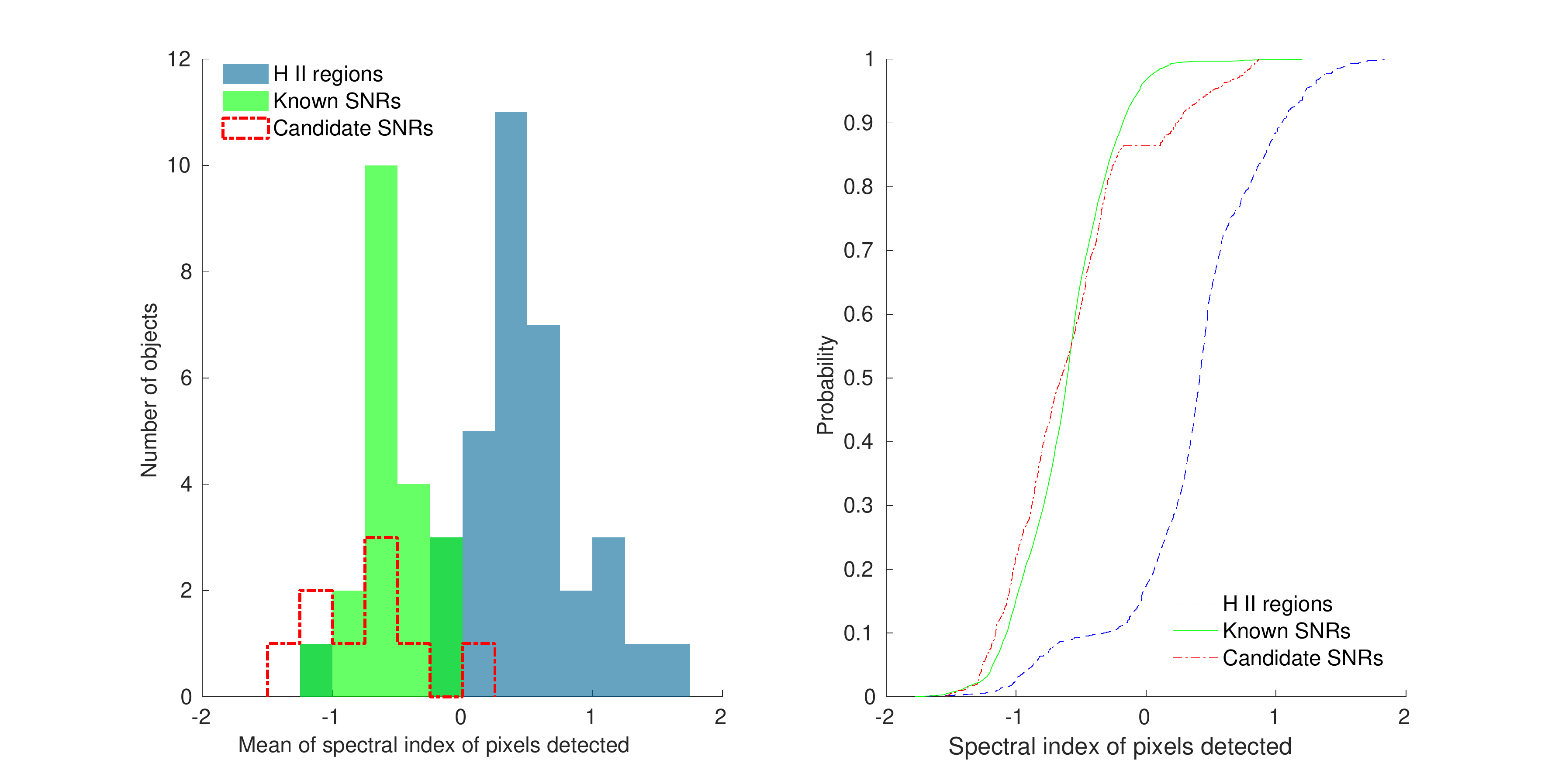}
  \caption{\edit1{\textit{Left:} Distribution of mean of spectral index values
  of pixels detected in both NVSS and TGSS, for the three samples.
  \textit{Right:} Cumulative probability distribution function of spectral
  index of the pixels belonging to the three samples. Objects containing
  background sources and H\begin{small}II\end{small}regions
  that overlap with known or candidate SNRs have been removed from the sample.}}
  \label{fig:spidx_hist_rem}
\end{figure*}

About 22\% of known SNRs, $\sim$60\% of SNR candidates and $\sim$78\% of
H\begin{small}II\end{small}regions are not detected in the TGSS. \edit2{The
TGSS is a snapshot survey with a large RMS noise of $\approx 25$ mJy beam$^{-1}$
in the Galactic plane after convolving to the beam size of NVSS. A pixel just
above detection limit in the NVSS data ($3\sigma = 1.35$ mJy beam$^{-1}$) will
not be detected in the TGSS unless it has a steep negative spectral index
($\sim -1.8$). A large fraction of SNRs, known and candidate, are not detected
in TGSS because they are not very bright at 1.4 GHz and do not have steep
negative spectral indices.}

\edit1{The distribution of spectral index of the three samples is shown in
Figure \ref{fig:spidx_hist_rem}. H\begin{small}II\end{small}regions containing
unrelated background sources or that overlap with possible or known SNRs have
been removed from the sample. We find one anomalous
H\begin{small}II\end{small}region (G050.317$-$00.421) that has
a large negative spectral index, consistent with the THOR continuum data
\citep{2016A&A...588A..97B}. This may be because of an AGN behind the
H\begin{small}II\end{small}region, or contamination from diffuse synchrotron
emission. SNRs that contain background sources were also removed from the
sample. As expected, there is a separation in the spectral index distribution
between thermal and non-thermal sources.}

\edit1{Only one candidate (G29.38+0.10) has a positive spectrum. It is
indicative of a PWN,} \edit2{but there exists no known pulsar along the
direction of the central emission \citep{2005AJ....129.1993M}.}

Non-thermal spectral index from the shell can confirm the status of the
candidates. Of the 76 candidates, only three have clear partial shell structure
with their spectral indices determined unambiguously: G27.06+0.04,
G51.21+0.11 and G53.41+0.03. \edit1{Several candidates such as G17.80$-$0.02 and
G36.68$-$0.14 have spectral indices from unidentified background sources. They
have been excluded from the discussion.}

\subsubsection{Candidate G27.06+0.04}

\begin{figure*}
  \gridline{\fig{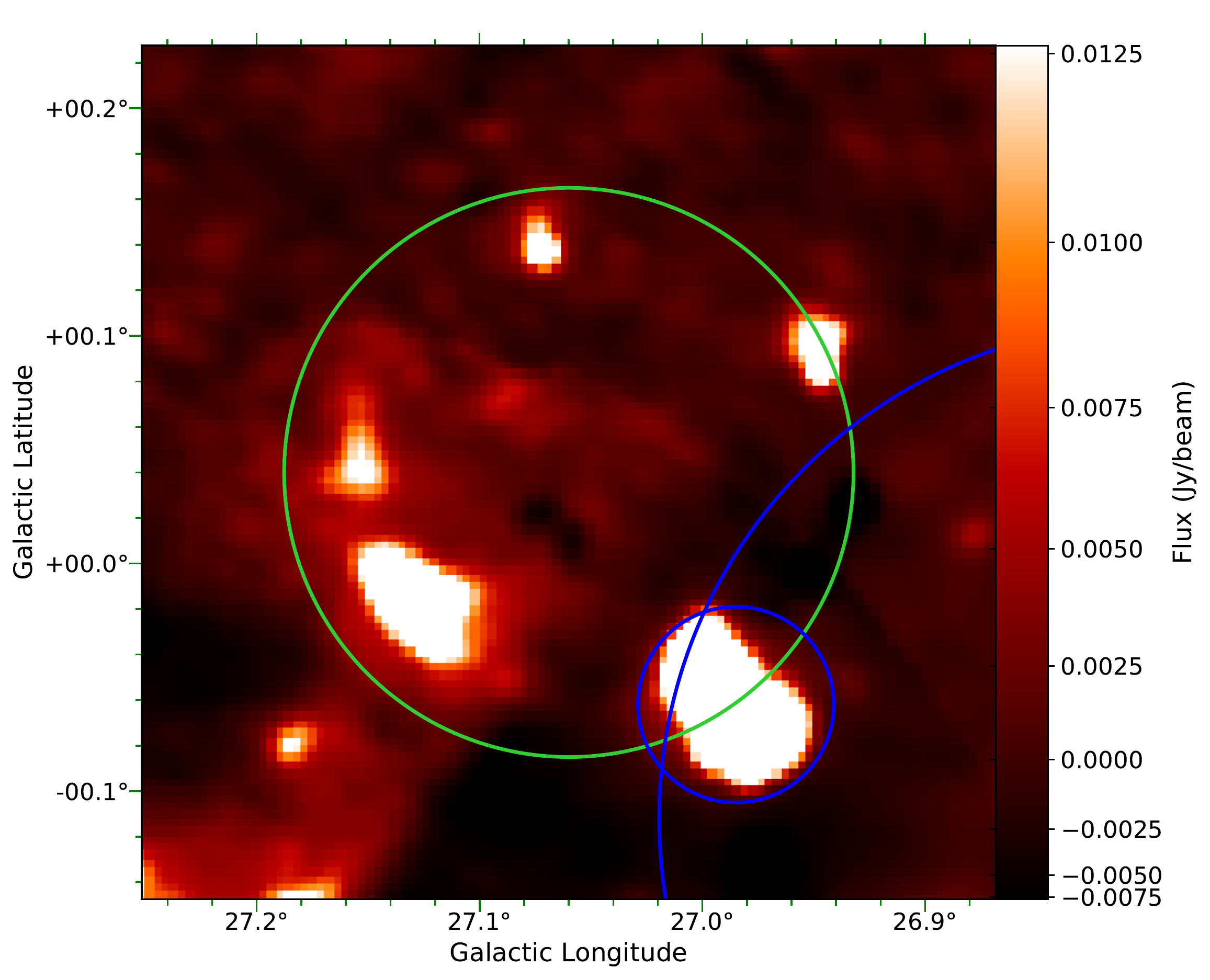}{0.5\textwidth}{(a)}
             \fig{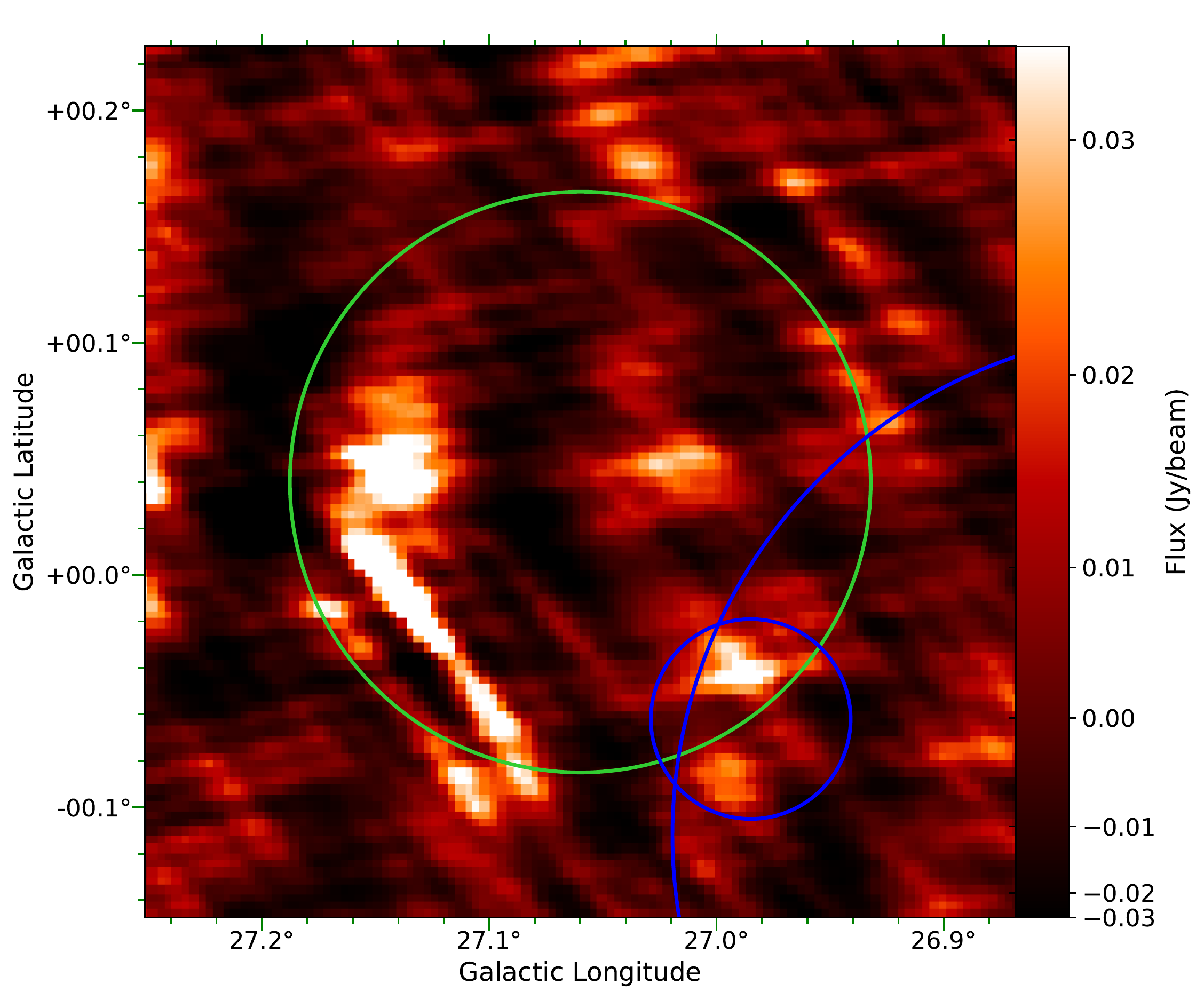}{0.5\textwidth}{(b)}
            }
  \gridline{\fig{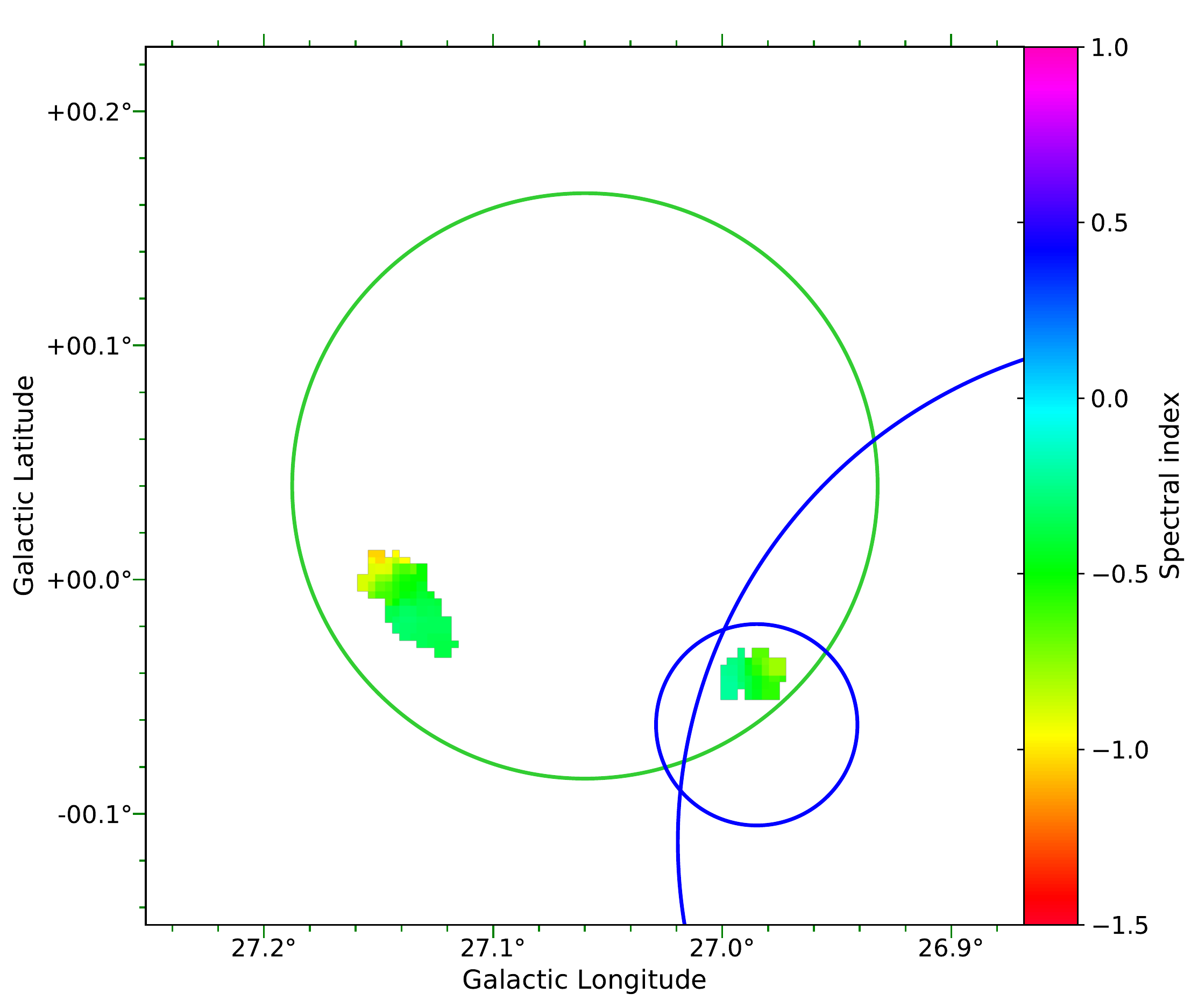}{0.5\textwidth}{(c)}
             \fig{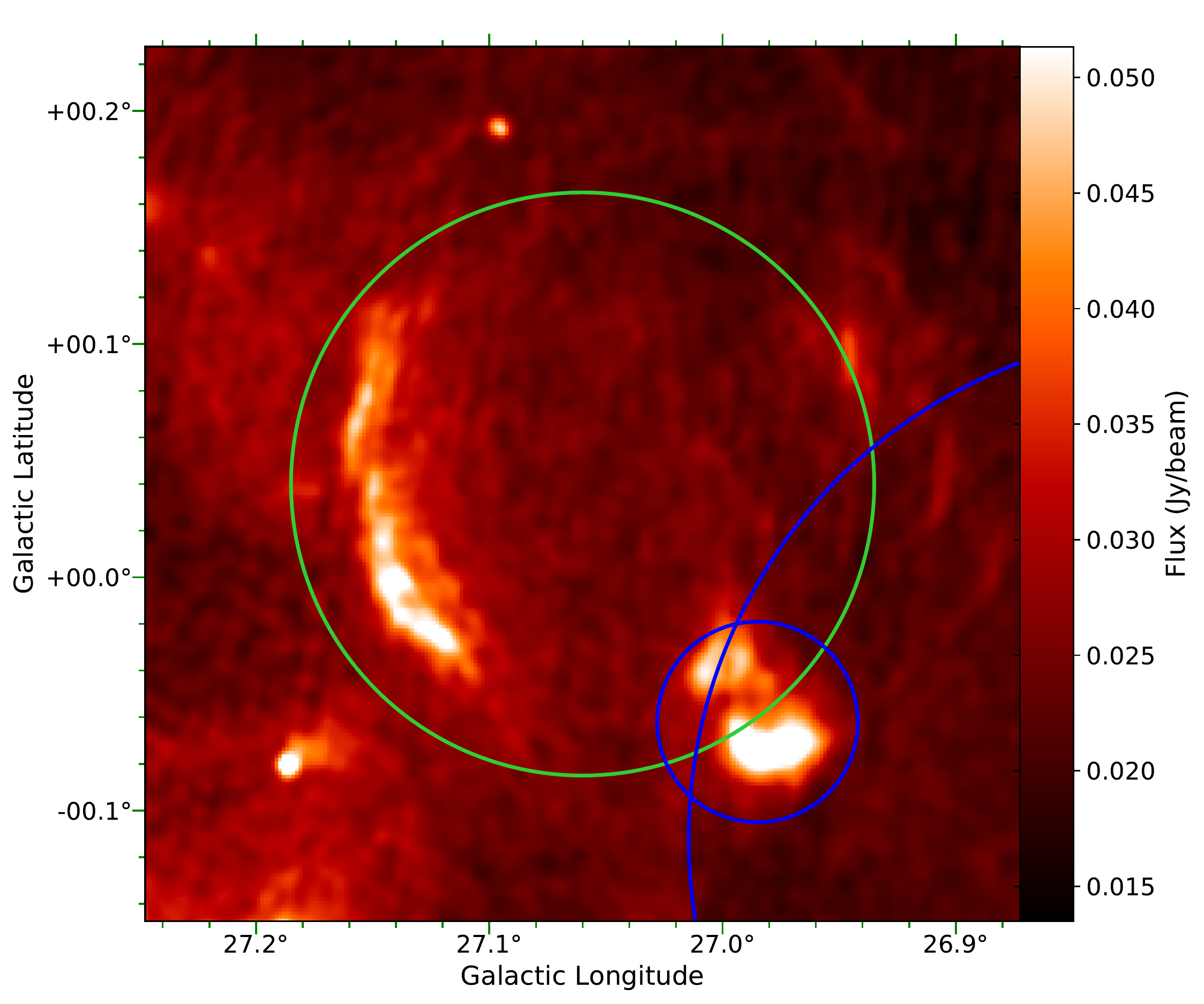}{0.5\textwidth}{(d)}
            }
  \caption{Candidate SNR G27.06+0.04: (a) NVSS 1.4 GHz Stokes $I$, (b)
  TGSS 150 MHz, (c) TGSS-NVSS spectral index map, (d) THOR+VGPS 1.4 GHz.
  \edit1{The candidate is enclosed by a green circle.
  H\begin{small}II\end{small}regions are marked by blue circles.}}
  \label{fig:cand_G27}
\end{figure*}

\edit1{Two regions of this candidate are detected in all three surveys ---
THOR, NVSS and TGSS (Figure \ref{fig:cand_G27}). The region to the south-west
overlaps with a bright H\begin{small}II\end{small}region
G026.984$-$00.062, but it shows a non-thermal spectrum as well. The other region
(eastern) has a partial shell morphology in THOR+VGPS data and its non-thermal
spectral index confirms its status as an SNR. A minor systematic gradient is
present on the spectral index map of its shell, even after accounting for errors
in the spectral index\footnote{available on
\url{http://tgssadr.strw.leidenuniv.nl/spidx/hips_spidxerr/} by
\citet{2018MNRAS.474.5008D}}. Similar gradients are observed in some other SNRs.
This could be inherent due to spatially varying properties such as optical
depth, magnetic field, leading to varied spectral index.}

\subsubsection{Candidate G51.21+0.11}

\begin{figure*}
  \gridline{\fig{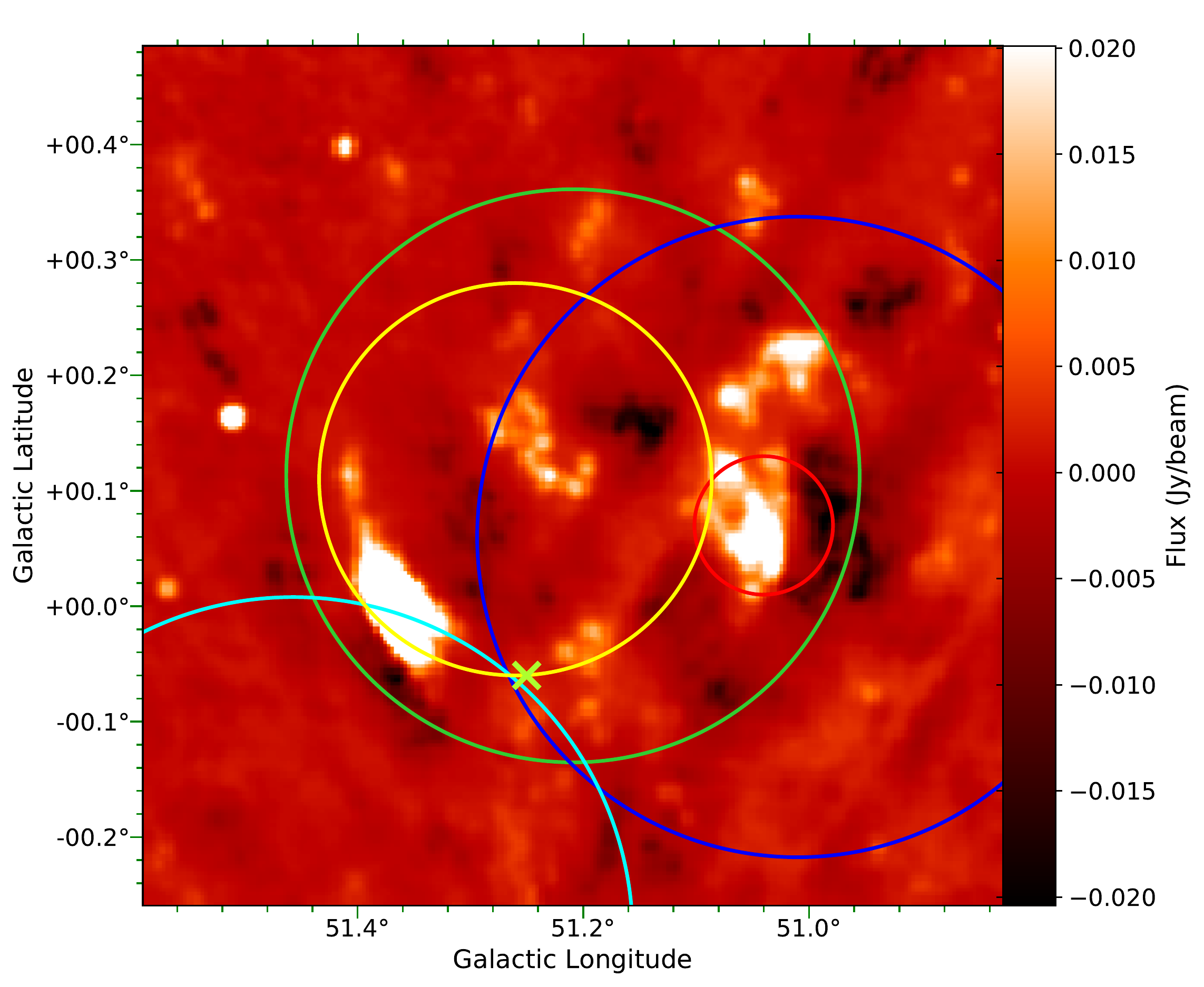}{0.5\textwidth}{(a)}
             \fig{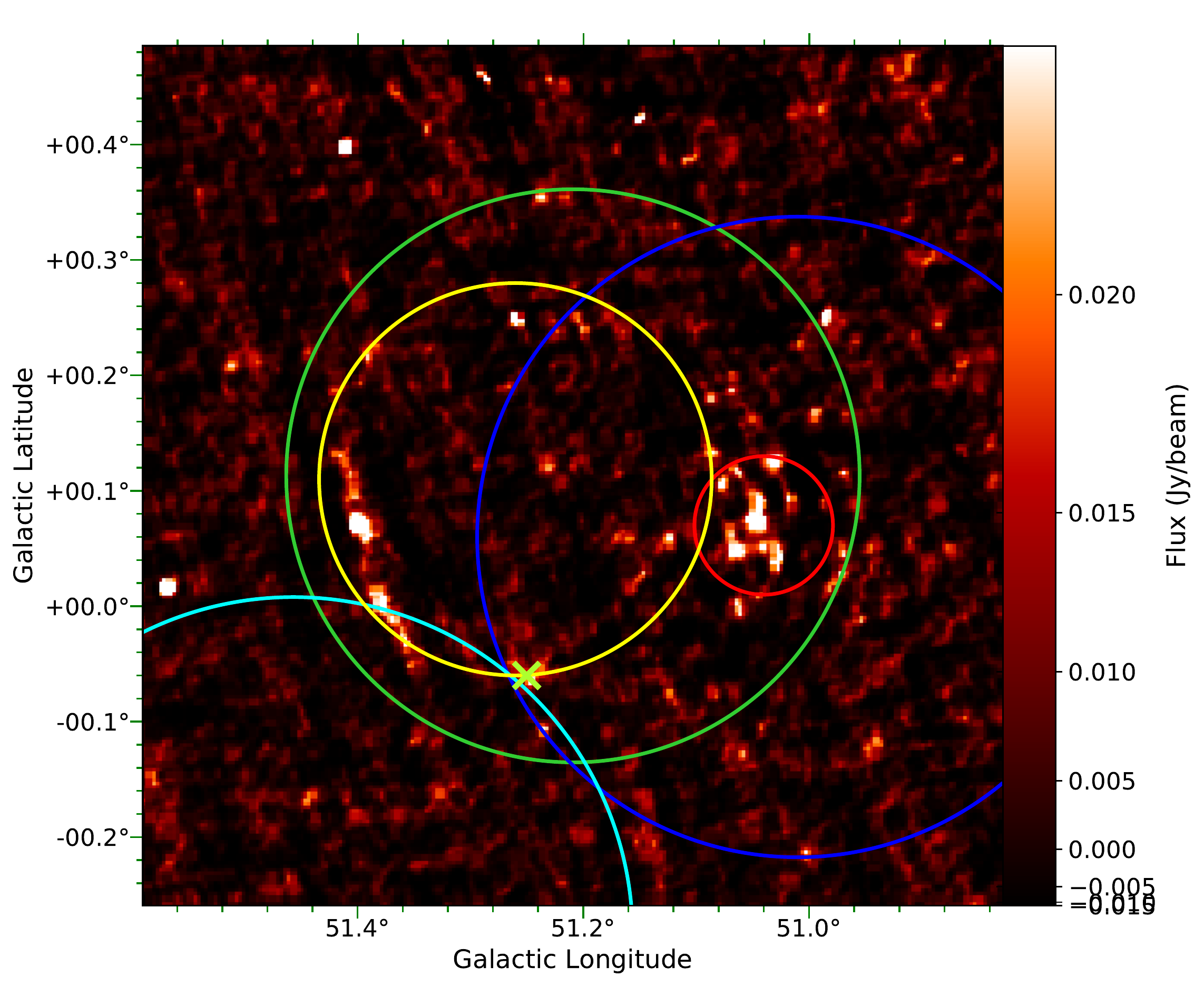}{0.5\textwidth}{(b)}
            }
  \gridline{\fig{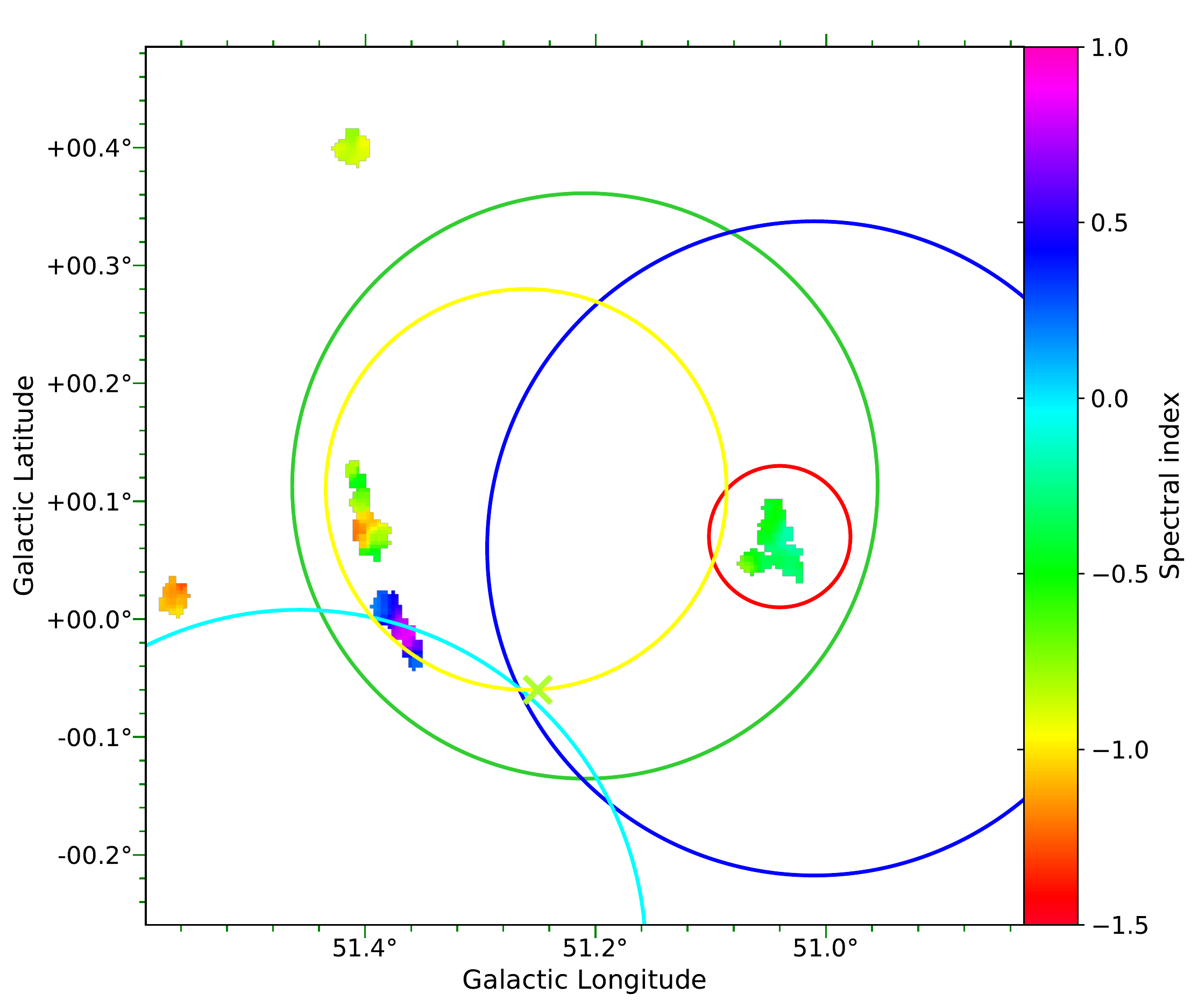}{0.5\textwidth}{(c)}
             \fig{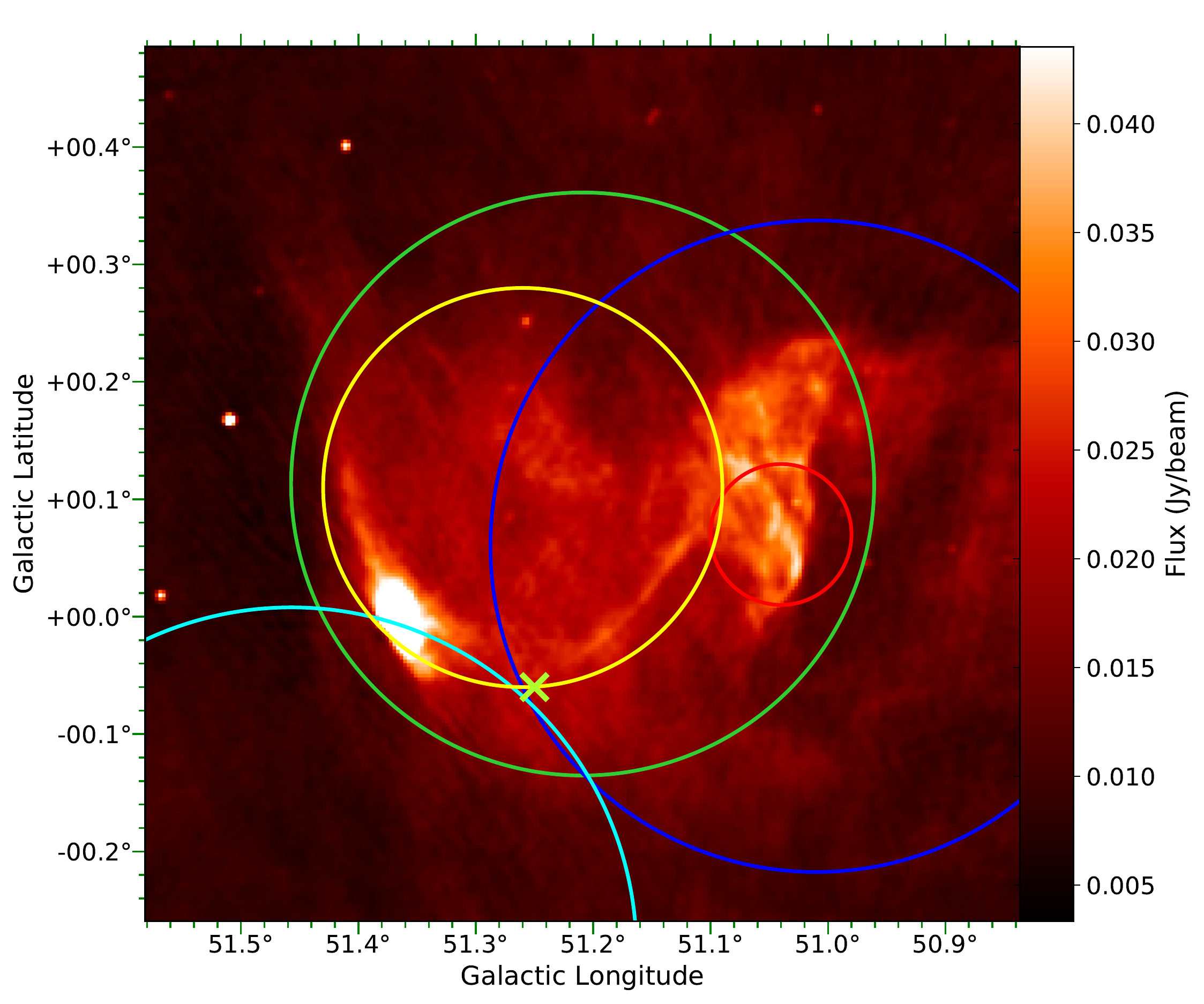}{0.5\textwidth}{(d)}
            }
  \caption{(a) NVSS 1.4 GHz Stokes $I$, (b) TGSS 150 MHz, (c) TGSS-NVSS spectral
  index map, (d) THOR+VGPS 1.4 GHz. \edit1{The extent
  of the following regions are marked by circles: candidate SNR G51.21+0.11
  (green), H\begin{small}II\end{small}region G051.010+00.060 (blue), candidate
  H\begin{small}II\end{small}region G051.457$-$00.286 (cyan), SNR G51.26+0.09
  (yellow) and SNR G51.04+0.07 (red). PSR J1926+1613 is marked with a cross.}}
  \label{fig:cand_G51}
\end{figure*}

\edit2{Candidate SNR G51.21+0.11 was observed by \citet{2017arXiv170608826D}
using the LOw Frequency ARray (LOFAR). They note that it has a morphology similar
to the one found by \citet{2017A&A...605A..58A}. We argue that this candidate is a
complex of two SNRs. The western region, named G51.04+0.07, was established by
\citet{2018arXiv180607452S} as a compact SNR. It is marked with a red circle
in Figure \ref{fig:cand_G51}.}

\edit2{A shell-type object centered at Galactic coordinates $l =
51.26\degree$, $b = 0.11\degree$ with radius $11.3'$ is visible in the THOR
data. It is marked with a yellow circle in Figure \ref{fig:cand_G51}. Thermal
emission is detected at $l = 51.38\degree$, $b = 0.00\degree$. This thermal emission has RRLs in
THOR data \citep{2016A&A...595A..32B} and strong 8.0 $\mu$m MIR emission from
GLIMPSE \citep{2003PASP..115..953B}. It is due to an overlap with the shell of
a candidate H\begin{small}II\end{small}region G051.457$-$00.286
\citep{2014yCat..22120001A} that is marked with a cyan circle in Figure
\ref{fig:cand_G51}.}

\edit2{A non-thermal spectrum is detected from the eastern part of the shell,
at $l = 51.40\degree$, $b = 0.08\degree$. The shell lacks MIR emission from all
regions other than the thermal emission mentioned above. The non-thermal
spectral index from a part of the shell confirms its nature. Hence the
originally defined candidate G51.21+0.11 is now re-classified as a complex of
two objects: compact SNR G51.04+0.07 \citep{2018arXiv180607452S} and shell type SNR G51.26+0.11. }

\subsubsection{Candidate G53.41+0.03}

\citet{2017arXiv170608826D} confirmed the status of candidate SNR G53.41+0.03
using observations from LOFAR and XMM-Newton. Our findings --- matching partial
shell shape in NVSS, TGSS and THOR data, and a non-thermal spectral index for
the shell --- agree with their \edit2{classification}.

\subsubsection{Other candidates with non-thermal spectra}
\label{other_cand}

\edit1{The SNR candidate \textit{G28.78$-$0.44} has extended emission in both
NVSS and TGSS data matching with the shell morphology in THOR data and
$\alpha \sim -0.75$. However, the spectral index was determined only for a
small part of the shell. It is possible that an AGN is the
cause of this spectral index. Hence we cannot confirm its status.}

\edit1{Compact SNR candidates \textit{G18.76$-$0.07}, \textit{G58.70$-$0.75} and
\textit{G59.68+1.25} have steep negative spectral indices, but we reserve
judgment on those because of lack of a shell morphology. We cannot rule out
AGNs as the cause of these spectral indices.}

\edit1{The possible shell of PWN G54.1+0.3 (candidate SNR \textit{G54.11+0.25})
is not visible in the TGSS data, consistent with the LOFAR observations of
\citet{2017arXiv170608826D}. We find only the PWN with $\alpha
\sim -0.25$. Hence, we cannot comment on the status of the candidate SNR
G54.11+0.25. \edit2{\citet{2017A&A...605A..58A} and \citet{2017arXiv170608826D}
discuss the possible shell around the PWN G54.1+0.3.}}

\subsubsection{Candidates with lower limits on spectral index}

\edit1{We note that if an object is detected in NVSS but not in TGSS, it does
not necessarily mean that the object is not an SNR, since the lower limit on the
spectral index for such pixels is $\sim-1.2$. Consider a typical pixel that is
detected in NVSS with a flux density 5 mJy beam$^{-1}$, but not in TGSS, which
implies that its flux density at 150 MHz cannot be greater than
3$\sigma_{TGSS}$\footnote{$\sigma_{TGSS} \approx 25$ mJy beam$^{-1}$ in the
Galactic plane after convolving TGSS images to the beam size of NVSS}. This
gives a lower limit: $\alpha > -1.2$. The shells of candidate SNRs G32.22$-$0.21
($\alpha > -0.84$) and G36.68$-$0.14 ($\alpha > -0.67$)\footnote{obtained from
the spectral index catalog --
\url{http://tgssadr.strw.leidenuniv.nl/doku.php?id=spidx\#catalog}} are two
examples. These limits in general are not helpful in identifying SNRs. We find
one SNR candidate (G28.56+0.00) with an interesting lower limit: $\alpha >
0.47$. This could be a misidentification by \citet{2017A&A...605A..58A}.}

\subsection{Polarization}

\begin{figure*}
  \centering
  \plotone{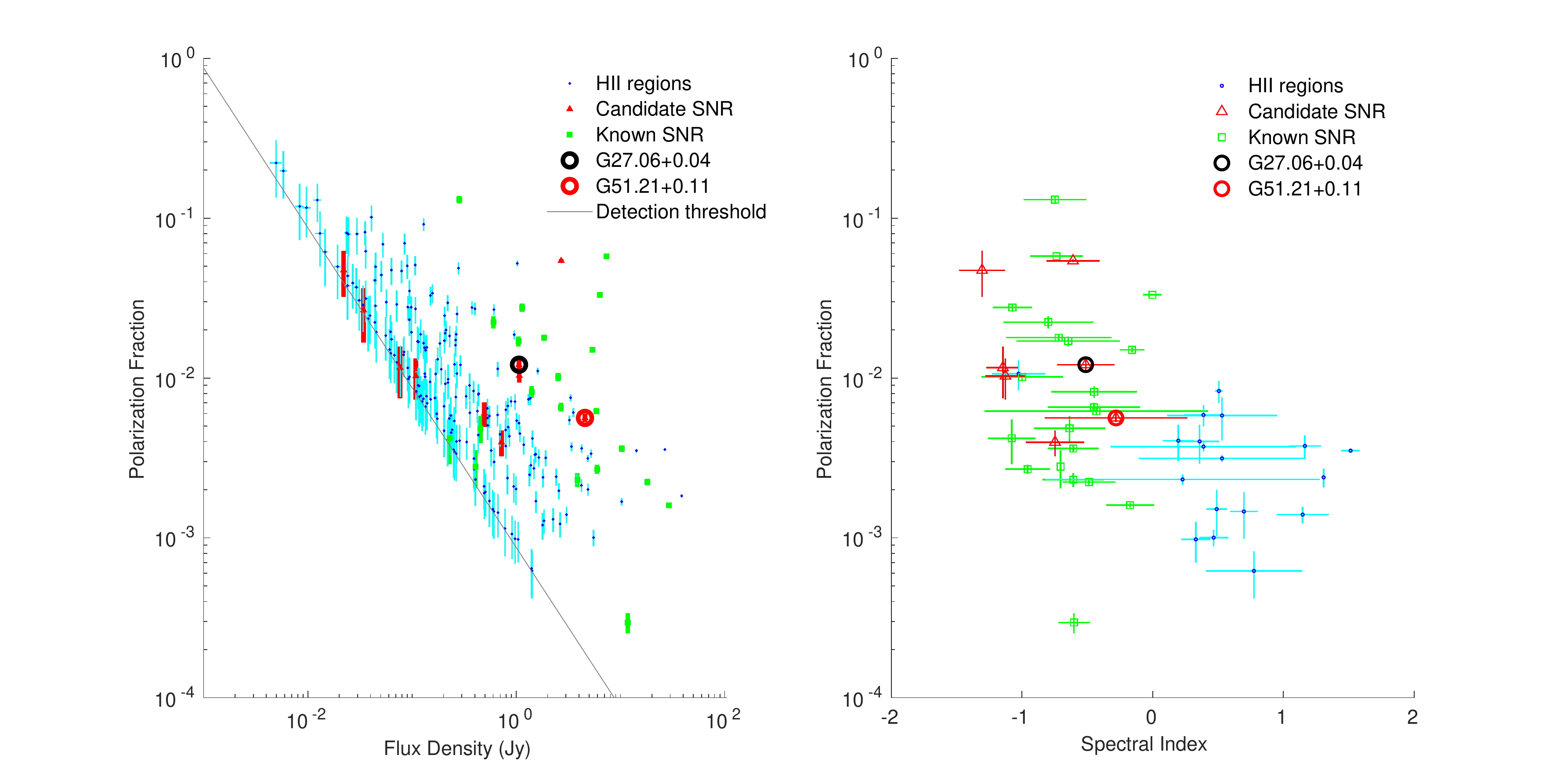}
  \caption{\textit{Left}: Fractional linear polarization against flux density
  from 1.4 GHz NVSS data. The detection threshold is the lower limit of
  polarization fraction that we can measure for a given flux density.
  Un-polarized sources are not shown. \textit{Right}: Fractional linear
  polarization against the mean of spectral index values detected.}
  \label{fig:P_FD}
\end{figure*}

\edit1{Fractional linear polarization is plotted against NVSS flux density and
spectral index for the three samples (Figure \ref{fig:P_FD}). Polarization is
higher for non-thermal sources as expected. We observe that there is no
pronounced offset that distinguishes SNRs from
H\begin{small}II\end{small}regions, despite SNRs having higher polarization in
general. This is due to polarization contamination. We do not attempt to correct
for polarization bias in this work. Estimating the noise in integrated Stokes Q
\& U maps of an extended source in the crowded Galactic plane is difficult.
Small-scale structure in the polarization angle of unrelated diffuse Galactic
emission acts as a non-Gaussian noise term that imposes its own bias in
polarized intensity, resulting in apparent polarized signal from
H\begin{small}II\end{small}regions. Some or all of the higher degree of
polarization of fainter objects may be the result of polarization bias.}

\subsection{Association with pulsars}\label{subsec:pulsar_assoc}

\edit2{Possible associations of pulsars with SNR candidates could be used to
argue for confirming the identification of SNR candidates. The following
conditions must be satisfied if a pulsar is associated with an SNR:
\begin{enumerate}[ {[}1{]} ]
\item the distance estimates to SNR and pulsar should be compatible,
\item the age estimates of SNR and pulsar should be similar,
\item the pulsar should be young enough to harbor a radio remnant (i.e. age
$\lesssim 100,000$ years) and
\item the transverse motion of the pulsar should be such that it could have
formed in the progenitor supernova explosion, and this transverse motion should
give a reasonable estimate of the pulsar kick velocity\footnote{Pulsar kicks arise from
asymmetry of the supernova explosion. The kick velocities typically range from
$200-500$ km s$^{-1}$ \citep[see][and references therein]{2000ASSL..254..127L}
but could be as high as $\sim 1100$km s$^{-1}$ \citep{2005ApJ...630L..61C}.}.
\end{enumerate}}

\edit2{We identify 15 pulsars along the line of sight of 13 candidates.
We do not have distance and age measurements of candidate SNRs, hence we
cannot check for conditions [1] and [2]. Only three pulsars (J1826$-$1256,
J1838$-$0537 and J1930+1852) satisfy condition [3], of which one pulsar
J1838$-$0537 has no distance measurement in the ATNF pulsar catalog
\citep{2005AJ....129.1993M}. }

\edit2{Pulsar J1826$-$1256 is along the line of sight of the candidate SNR
G18.45$-$0.42. It is associated with a gamma ray source
\citep{2012ApJS..199...31N} and an X-Ray source \citep{2001ApJS..133..451R}.
However, compared to the angular size of the candidate, the associated high
energy sources are too small to make a meaningful claim for an association.
Given a distance of 1.55 kpc and an age of 14,400 years for the pulsar J1826$-$1256
\citep{2005AJ....129.1993M}, we find that the transverse kick velocity should
be $\sim$210 km s$^{-1}$ if the candidate SNR G18.45$-$0.42 and the pulsar are
indeed the results of the same supernova.}

\edit2{Pulsar J1930+1852 along the line of sight of candidate SNR G54.11+0.25
cannot be used to confirm the nature of the candidate due to its ambiguous
shell \citep[see: Section \ref{other_cand};][]{2017A&A...605A..58A,2017arXiv170608826D}.}

\section{Conclusions and Future Work}

We have shown that the statistics of SNR candidates follows the sample of known
SNRs more closely than that of H\begin{small}II\end{small}regions in spectral
index and linear polarization. \edit2{However, the fractional polarization
could not be used to discriminate between SNRs and H\begin{small}II\end{small}regions
because of contamination by diffuse polarized emission in the Galactic
Plane.} Compact sources and overlaps with known or candidate SNRs account for
most of the steep negative spectra in H\begin{small}II\end{small}regions.
There is only one H\begin{small}II\end{small}region G050.317$-$00.421 with
an apparent non-thermal spectrum that needs to be explained.

Despite the above shortcomings, spectral index data, along with morphology,
confirmed the status of G27.06+0.04 and G51.26+0.11 as SNRs. \edit1{There are
three other candidate SNRs with non-thermal spectral indices (G18.76$-$0.07,
G58.70$-$0.75 and G59.68+1.25) but no shell morphology. High energy emissions
and a high degree of polarization might confirm their nature. Ongoing
work on the THOR survey includes a careful analysis of the polarization data
(Stil et al. in prep). Though THOR data
are not ideal for deriving the spectra of large structures, they work well for
compact sources \citep[][Wang et al. submitted]{2016A&A...588A..97B}. Candidate
SNRs G28.56+0.00 and G47.15+0.73 are such sources (angular size $< 3'$). They
are detected in 1.4 GHz NVSS but not in 150 MHz TGSS. Future spectral index
information from THOR could be used to study these candidates.} \edit2{Using
optical emission lines could help to distinguish SNRs since they have elevated
values for [S II]:H$\alpha$ compared with H\begin{small}II\end{small}regions
and the lines are often broader \citep{2018ApJ...855..140L}.}

\edit2{We have been able to confirm the identification of only two candidates
out of 76 using spectral index and morphology. Several candidates have not
been detected in TGSS, some not in NVSS as well. Both these
data are from snapshot surveys, which are not well suited to study low surface
brightness emissions. On the other hand, compact candidate SNRs -- despite
favorable spectral index measurements -- could not be confirmed because of
confusion with background sources (AGNs). The low rate in confirming
the identification of candidate SNRs underlines the importance of future
Galactic plane surveys with better sensitivity and high angular resolution.}

\edit2{We could not find any pulsar associations with candidate SNRs. More data
on proper motions of pulsars, age and distance measurements of candidate SNRs
can be used to argue for or against an association.} \edit1{Proper motions of
pulsars can be measured by comparing their current positions with the positions
in the ATNF pulsar catalog. Distances to pulsars can be estimated from
dispersion measure and an electron density model \citep{2017ApJ...835...29Y},
or through the \textit{kinematic} method. Astrometric observations by Very
Long Baseline Array also can be used to measure parallaxes and proper motions
of pulsars \citep{2009ApJ...698..250C}. }

\acknowledgements

R.D. is a recipient of INSPIRE scholarship from the Department of Science and
Technology, Government of India. N.R. acknowledges support from the Infosys
Foundation through the Infosys Young Investigator grant. H.B. and Y.W.
acknowledge support from the European Research Council under the Horizon 2020
Framework Program via the ERC Consolidator Grant CSF-648505. This research has
made use of NASA's Astrophysics Data System, data from surveys by VLA (run by
the National Radio Astronomy Observatory) and GMRT (run by the National Centre
for Radio Astrophysics of the Tata Institute of Fundamental Research).
The National Radio Astronomy Observatory is a facility of the National Science
Foundation operated under cooperative agreement by Associated Universities, Inc.

\software{ALADIN interactive sky atlas \citep{2000A&AS..143...33B},
          APLpy \citep{2012ascl.soft08017R},
          SAOImage DS9 \citep{2003ASPC..295..489J}\footnote{\url{http://ds9.si.edu/site/Home.html}}}

\end{document}